\documentclass[oldversion,longauth]{aa}
\usepackage{aalongtable}
\usepackage{natbib}
\usepackage{txfonts}
\usepackage[english]{babel}
\usepackage{graphicx}
\usepackage{tabularx}
\usepackage{array}
\usepackage{amssymb}
\usepackage{color}
\usepackage{soul}

\def\sun{\hbox{$\odot$}}

\def\kms{${\rm km\,s}^{-1}$}
\def\ergs{${\rm erg\,s}^{-1}$}

\begin{document}
\title{Eddington ratios of faint AGN at intermediate redshift: Evidence for a
population of half-starved black holes
  \thanks{Based on data obtained with the European Southern Observatory
  Very Large Telescope, Paranal, Chile, program 070.A-9007(A),
  272.A-5047, 076.A-0808, and partially on data obtained at the
  Canada-France-Hawaii  Telescope.}}

\author{
I. Gavignaud \inst{1}
\and L. Wisotzki    \inst{1}
\and A. Bongiorno \inst{2}
\and S. Paltani \inst{3,4}
\and G. Zamorani \inst{5} 
\and P. M\o ller\inst{6}
\and V. Le Brun \inst{7}
\and B. Husemann \inst{1}
\and F. Lamareille \inst{8}
\and M. Schramm \inst{1}
\and O. Le F\`evre \inst{7}
\and D. Bottini \inst{9}
\and B. Garilli \inst{9}
\and D. Maccagni \inst{9}
\and R. Scaramella \inst{10,11}
\and M. Scodeggio \inst{9}
\and L. Tresse \inst{7}
\and G. Vettolani \inst{10}
\and A. Zanichelli \inst{10}
\and C. Adami \inst{7}
\and M. Arnaboldi \inst{12}
\and S. Arnouts \inst{7}
\and S. Bardelli  \inst{5}
\and M. Bolzonella  \inst{5} 
\and A. Cappi    \inst{5}
\and S. Charlot \inst{13}
\and P. Ciliegi    \inst{5}  
\and T. Contini \inst{8}
\and S. Foucaud \inst{14}
\and P. Franzetti \inst{9}
\and L. Guzzo \inst{15}
\and O. Ilbert \inst{16}
\and A. Iovino \inst{15}
\and H.J. McCracken \inst{13,17}
\and B. Marano \inst{18}  
\and C. Marinoni \inst{8}
\and A. Mazure \inst{8}
\and B. Meneux \inst{9,15}
\and R. Merighi   \inst{5} 
\and R. Pell\`o \inst{8}
\and A. Pollo \inst{8}
\and L. Pozzetti    \inst{5} 
\and M. Radovich \inst{12}
\and E. Zucca    \inst{5}
\and M. Bondi \inst{10}
\and G. Busarello \inst{12}
\and O. Cucciati \inst{15,18}
\and S. de la Torre \inst{8}
\and L. Gregorini \inst{10}
\and Y. Mellier \inst{13,17}
\and P. Merluzzi \inst{12}
\and V. Ripepi \inst{12}
\and D. Rizzo \inst{2}
\and D. Vergani \inst{9}
}

\institute{
Astrophysikalisches Institut Potsdam, An der Sternwarte 16, D-14482 Potsdam,
Germany
\and
Max-Planck-Institut f\"ur Extraterrestrische Physik, Giessenbachstr., D-85741,
Garching, Germany
\and
Integral Science Data Centre, ch. d'\'Ecogia 16, CH-1290 Versoix
\and
Geneva Observatory, ch. des Maillettes 51, CH-1290 Sauverny
\and
INAF-Osservatorio Astronomico di Bologna - Via Ranzani,1, I-40127, Bologna, Italy
\and
European Southern Observatory, Karl-Schwarzschild-Strasse 2, D-85748
Garching bei M\"unchen, Germany
\and
Laboratoire d'Astrophysique de Marseille (UMR6110), CNRS-Universit\'e de
Provence, 38 rue Frederic Joliot-Curie, F-13388 Marseille Cedex 13
\and
Laboratoire d'Astrophysique de Toulouse-Tarbes, Universit\'e de Toulouse,
CNRS, 14 avenue Edouard Belin, F-31400 Toulouse, France.
\and
IASF-INAF - via Bassini 15, I-20133, Milano, Italy
\and
IRA-INAF - Via Gobetti,101, I-40129, Bologna, Italy
\and
INAF-Osservatorio Astronomico di Roma - Via di Frascati 33,
I-00040, Monte Porzio Catone,
Italy
\and
INAF-Osservatorio Astronomico di Capodimonte - Via Moiariello 16, I-80131, Napoli,
Italy
\and
Institut d'Astrophysique de Paris, UMR 7095, 98 bis Bvd Arago, 75014
Paris, France
\and
School of Physics \& Astronomy, University of Nottingham,
University Park, Nottingham, NG72RD, UK
INAF-Osservatorio Astronomico di Brera - Via Brera 28, Milan,
Italy
\and
Canada France Hawaii Telescope corporation, Mamalahoa Hwy,  
Kamuela, HI-96743, USA
\and
Observatoire de Paris, LERMA, 61 Avenue de l'Observatoire, 75014 Paris, 
France
\and
Universit\`a di Bologna, Dipartimento di Astronomia - Via Ranzani,1,
I-40127, Bologna, Italy
\and
Universit\'a di Milano-Bicocca, Dipartimento di Fisica - 
Piazza delle Scienze, 3, I-20126 Milano, Italy
}

\offprints{I. Gavignaud, \email{igavignaud~@aip.de}}

\date{Received ..., ....; accepted ..., ....}

\abstract{We use one of the deepest spectroscopic samples of broad line active
      galactic nuclei (AGN) currently available, extracted from the VIMOS VLT
      Deep Survey (VVDS), to compute \ion{Mg}{ii} and \ion{C}{iv} virial
      masses estimate of 120 super-massive black holes in the redshift range
      $1.0<z<1.9$ and $2.6<z<4.3$.
      We find that the mass-luminosity relation shows considerably
      enhanced dispersion towards low AGN luminosities ($\log L_\mathrm{bol}
      \sim 45$).  
      At these luminosities, there is a substantial fraction of black holes
      accreting far below their Eddington limit
      ($L_\mathrm{bol}/L_\mathrm{Edd} < 0.1$), in marked contrast to what is
      generally found for AGN of higher luminosities. We speculate that these
      may be AGN on the decaying branch of their light-curves, well past their
      peak activity.  This would agree with recent theoretical
      predictions of AGN evolution.  

      In the electronic Appendix of this paper we publish an update of the VVDS
        type-1 AGN sample, including the first and most of the second epoch
        observations. This sample contains 298 objects of which 168 are new.
}

\authorrunning{Gavignaud et al.}
\titlerunning{Eddington ratios of faint AGN at intermediate redshift}

\maketitle

\section{Introduction}

The mass scaling relations of super-massive black holes in present-day galaxies
\citep[e.g.,][]{Gebhardt2000,Ferrarese2000} imply that black hole growth must
be 
closely connected to the overall formation and evolution of galaxies. Most
of the mass locked up in black holes today was probably accumulated through
accretion in discrete phases of nuclear activity, as suggested by the
consistency between the estimate of the black hole mass density at
$z\approx 0$ and that derived from the integrated AGN luminosity density
\citep{Soltan1982,Yu2002,Marconi2004}.  

Accretion histories of individual black holes are essentially unconstrained
from observations. By looking at AGN one may at least catch snapshots of the
black hole growth process, especially when black hole masses and thus
accretion rates can be estimated. There has been significant progress in this
direction over the last years, and it has been demonstrated that single-epoch
spectroscopic and photometric measurements of AGN with broad emission lines
(type~1 AGN) allow one to estimate black hole masses to an accuracy of
the order of $\pm 0.5$~dex \citep{Vestergaard2002,McLure2002,Collin2006}. With
this approach it has been possible to explore the distribution of Eddington
ratios for large AGN surveys \citep{McLure2004,Kollmeier2006}. 

These studies have shown that powerful type~1 AGN appear to accrete at rates
close to the Eddington limit with remarkable uniformity. Yet, periods of
activity must be followed by a transition from high-luminosity
near-Eddington states to almost quiescent black holes.  
Unless this transition is
rather abrupt, there should therefore also be a population of AGN with
significantly lower Eddington ratios, but still recognizable as bona-fide
AGN. In this paper we report on observations of such a population at
intermediate redshifts, based on black hole mass estimates that we derive for
a new sample of faint AGN with complete spectroscopic identification.

In this work, absolute luminosities are computed assuming a flat universe
  with cosmological parameters $\Omega_m=0.3$, $\Omega_\Lambda=0.7$ and
  $H_0=70$\kms.

\section{The sample}

The VVDS (VLT-VIMOS Deep Survey) is a purely $I$-band flux limited
spectroscopic survey designed to study the evolution of galaxies, AGN, and
large scale structure. It comprises two subsets: a `deep' survey with a limit
of $I_{AB} \le 24$ \citep{LeFevre2005} and a `wide' and a shallower
survey with $I_{AB} \le 22.5$ \citep{Garilli2008}. Both surveys utilize the
VIMOS multi-object spectrograph on the ESO-VLT to take spectra of objects
above the flux limit, irrespective of their morphological properties or
colors, albeit with sparse target sampling rate (for details see the above
mentioned papers).

About 1~\% of all VVDS targets can be classified as type 1 AGN on the basis of
their broad emission lines. 
From the VVDS we can therefore construct AGN samples that have two
advantages over most other surveys: (i) 
The very faint limiting magnitude, which is even deeper than that of the
multi-color photometric COMBO-17 survey \citep{wolf2003}; and (ii) the simple
selection criterion, which requires only the presence of a broad emission line
($FWHM \ge 1000$\kms) in any given spectrum.  This way, we have recently
constructed a well-defined sample of type1 AGN which is described in detail by
\citet{Gavignaud2006}. In that paper we demonstrated that, since the sample is
unaffected by morphological or color pre-selection biases, it is also much
less prone to incompleteness due to host galaxy contamination.  The sample has
already been used to investigate the AGN luminosity function and its evolution
\citep{Bongiorno2007}. Here we exploit the
spectroscopic properties of that sample, containing 130 broad-line AGN,
 supplemented by 168 AGN of the VVDS second epoch data. 
The merged updated catalog of AGN is published in Appendix \ref{epoch2} of 
this paper.
It contains 222 and 76 AGN from the `wide' and the `deep' survey respectively. 
The median redshift is $z\sim 1.8$ (roughly equal for the wide and deep
subsets), with a broad distribution of redshifts within $1 \la z \la 3 $.

\section{Black hole masses and Eddington ratios}\label{sec:bhm}

\begin{figure}
  \begin{center}
    \includegraphics[width=0.95\linewidth]{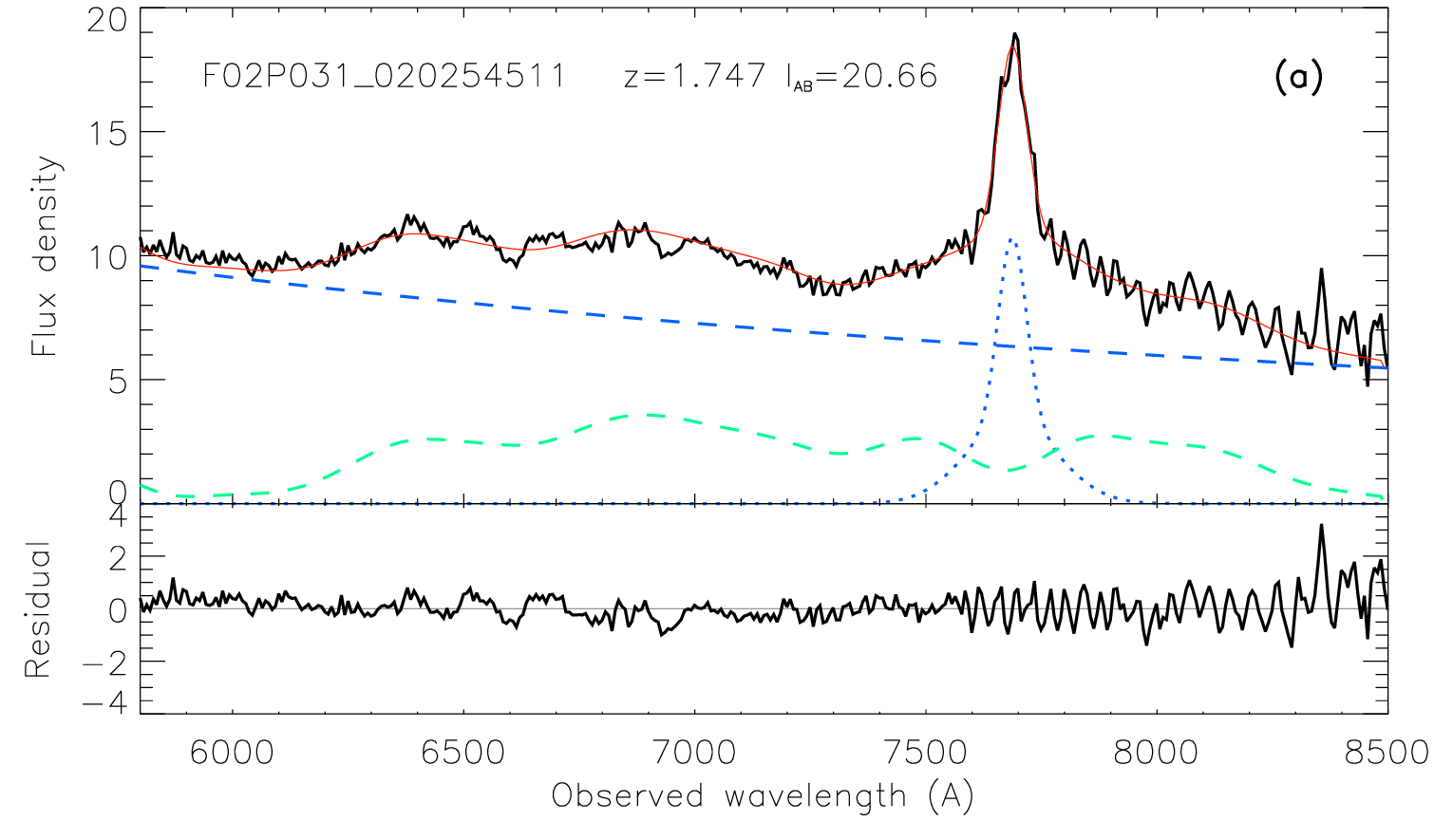} 
    \includegraphics[width=0.95\linewidth]{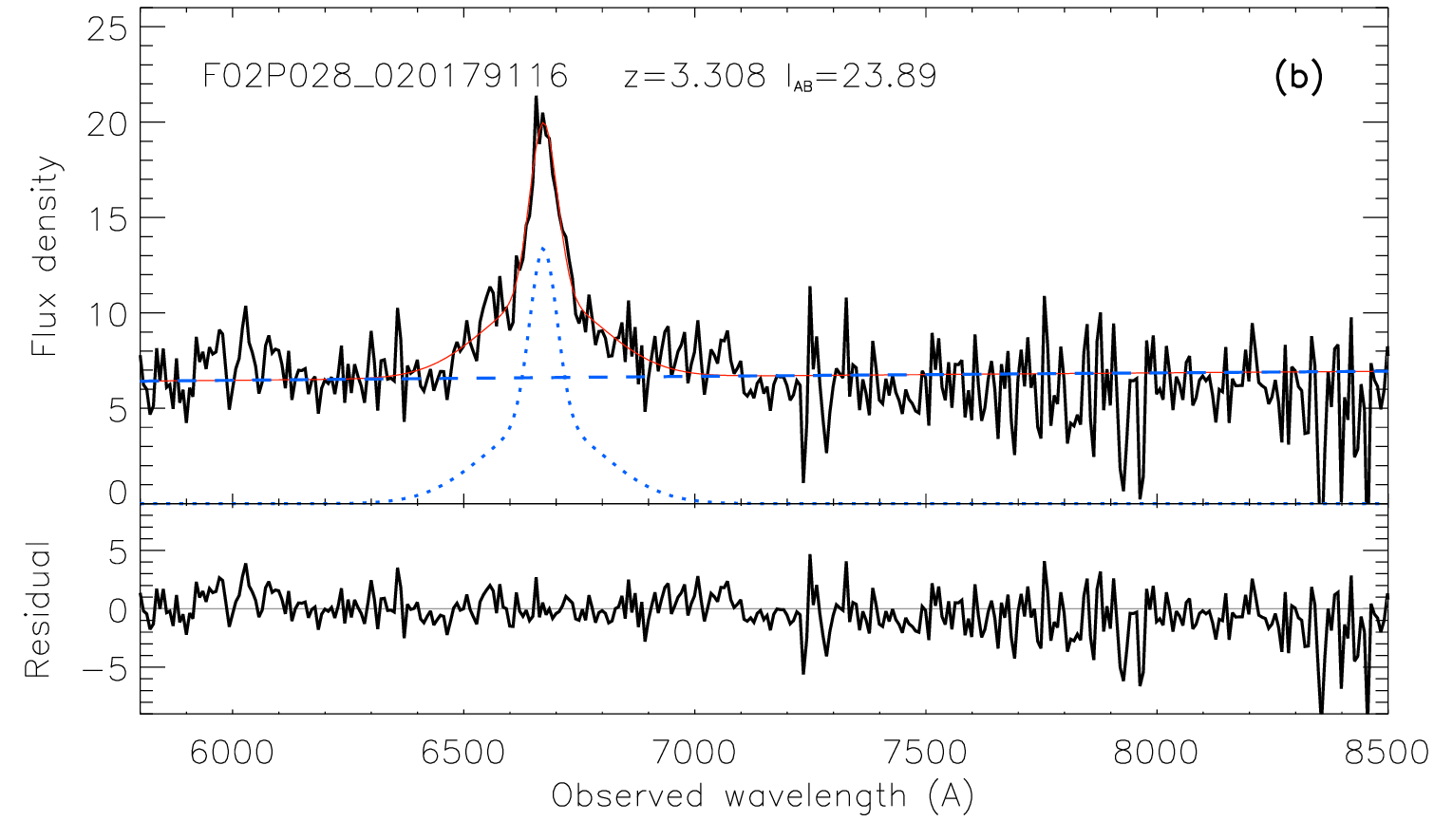} 
\caption[Examples of spectra]
{\label{fig:spec}
Two examples of emission line fits to the spectra. In (a) we show an object from our
  low-redshift sample with the \ion{Mg}{ii} line and in (b) an
  example of the high-redshift sample with the \ion{C}{iv} line.
The observed spectra are displayed in black, the fits are overplotted in red.
Each fit is a combination of a power-law continuum (blue dashed line), a
double-Gaussian model of the broad-emission line (blue dotted line) and,
only for the \ion{Mg}{ii} sample, a broadened empirical template of the
\ion{Fe}{ii} pseudo-continuum emission (green dashed-line).
}
\end{center}
\end{figure}

In order to estimate the black hole masses in type~1 AGN from single
epoch spectroscopy, it must be assumed that the line-emitting `clouds' are 
roughly in virial equilibrium, and that the size of the broad-line region 
(BLR) is closely correlated with the luminosity of the AGN. 
The black hole mass is then given by the virial relation \citep{Collin2006},
$M_\mathrm{BH} = f\,(R\,\Delta V^2)/\mathcal{G}$,
where $\mathcal{G}$ is the gravitational constant, $R$ is the size of the BLR,
which in turn is estimated from the continuum luminosity, 
$f$ is a dimensionless factor close to unity which reflects the
unknown geometry and inclination of the BLR, and $\Delta V$ represents the
velocity broadening of a given broad emission line. $\Delta V$ can be
  estimated using either the line FWHM or the line velocity dispersion
$\sigma_\mathrm{l}$.

We have applied this approach to our sample of 298 type~1 AGN.  The
spectral range available for measuring line widths is 5700~\AA--8200~\AA .
Consequently, for $1.0 \la z \la 1.9$, the spectra contain the \ion{Mg}{ii}
$\lambda$2798 emission line, while \ion{C}{iv} $\lambda$1550 is accessible for
$2.6 \la z \la 4.3$.
 
In the case of \ion{Mg}{ii}, we applied an iterative procedure to subtract
the \ion{Fe}{II} contamination from the AGN continuum using a template
kindly provided by M.~Vestergaard \cite[see][]{Vestergaard2001}. Since this
template is derived from 
the observed spectrum of a narrow line Seyfert I, it is difficult to deblend
the \ion{Fe}{ii} pseudo-continuum emission from other emission lines. In particular, the
template contains no flux under the \ion{Mg}{ii} line itself although
some amount of flux is expected from theoretical models \citep{Sigut2003}.
The effect of adding flux to the empirical template has been recently
quantified by \cite{Fine2008} 
and is found to be negligible in view of the other errors.
In this work we used the unmodified template.

The \ion{Mg}{ii} and  \ion{C}{iv} emission line profiles were modeled by
  a superposition of two Gaussian components; the line widths were 
  obtained from these fits.
The measurements were then corrected for the finite spectrograph resolution
assuming that $\Delta\lambda_\mathrm{obs}^2 =
\Delta\lambda_\mathrm{intrinsic}^2+\Delta\lambda_\mathrm{res}^2$.  The mean
instrumental resolution of the VVDS spectra corresponds to
$\Delta\sigma_\mathrm{res} = 350$~\kms. 
Errors on the velocity measurements are obtained by combining the nominal
errors of the fit parameters and the uncertainties due to the adopted
continuum level. 
Fig. 1 shows two examples of fits to the spectra (continuum + emission
  lines) representative for the two redshift intervals.

Objects with a mean signal-to-noise (S/N) ratio per pixel lower than 7, in the
vicinity of the emission line, were excluded from further analysis 
(this concerns 8 \ion{Mg}{ii} and 8 \ion{C}{iv} objects).
20 of the high-redshift \ion{C}{iv} line profiles and 4 of the
low-redshift \ion{Mg}{ii} were 
heavily affected by associated absorption or instrumental problems, 
and these were also eliminated.  
After these cuts we remained with a sample of 120
objects, 91 of which feature the \ion{Mg}{ii}, and 29 of which
feature the \ion{C}{iv} line. The median redshift is 1.5 for the \ion{Mg}{ii}
  subsample and 3.1 for the \ion{C}{iv} subsample, respectively.

We flux-calibrated our spectra
by scaling them to the $I$ band photometry
in the CFHT images used as input to the VVDS.
Monochromatic luminosities at given rest-frame
wavelengths were then directly measured from the spectra.

In order to apply the virial relation to measurements of the \ion{Mg}{ii}
emission line we used the empirical calibration by \citet{McLure2004}
\begin{equation}
\log{\frac{M_{BH}}{M\sun}}  = 
\log{\left(\mathrm{FWHM}_{1000}^2\,((\lambda L)_{44,3000})^{0.62}\right)} + 6.51 
\label{eq:Mbh_MgII}
\end{equation}
where $\mathrm{FWHM}_{1000}$ is the FWHM of the line in units of
1000~\kms, and $\lambda L_\mathrm{44,3000}$ is the monochromatic luminosity 
at $\lambda = 3000$~\AA, expressed in units of $10^{44}$~erg~s$^{-1}$. 

For AGN where only \ion{C}{iv} could be measured,
we employed the recent relation by 
\cite{Vestergaard2006},
\begin{equation}
\log{\frac{M_{BH}}{M\sun}}  = \log{\left(\sigma_{1000}^2((\lambda
    L)_\mathrm{44,1350})^{0.53}\right)} + 6.73  
\label{eq:Mbh_CIV}
\end{equation}
where $\sigma_{1000}$ is the emission line velocity dispersion in units
of 1000 \kms and $(\lambda L)_{44,1350}$ is the monochromatic luminosity at
1350 \AA, expressed in units of $10^{44}$~erg~s$^{-1}$.

Bolometric luminosities were derived from the monochromatic
ones, multiplied by a correction factor $f_\mathrm{bol}$.
It is now established that, on average, at UV and optical wavelengths this
correction factor increases towards lower luminosities
\citep[e.g.][]{Richards2006,Steffen2006}. 
\cite{Hopkins2007} provide an empirical model of AGN SED which varies with
bolometric luminosity and is calibrated from a large number of observational
studies\footnote{see http://www.cfa.harvard.edu/\~{}phopkins/Site/qlf.html and
references therein}.
Following this model, $f_\mathrm{bol}(3000\mbox{\AA})$ decreases
from 6.8 to 5.6  over the luminosity range $\log L_\mathrm{bol} = [44.8,46.2]$, 
while $f_\mathrm{bol}(1350\mbox{\AA})$ varies between 4.2 and 3.7 for 
$\log L_\mathrm{bol} = [45.2,46.4]$. 

Together with black hole masses and bolometric luminosities we also estimated
the dimensionless `Eddington ratios' $\epsilon =
L_\mathrm{bol}/L_\mathrm{Edd}$, where $L_\mathrm{Edd}$ is the Eddington
luminosity of the black hole assuming spherically symmetric accretion.

\section{Results}

\begin{figure*}
  \begin{center}
    \includegraphics[width=0.95\linewidth]{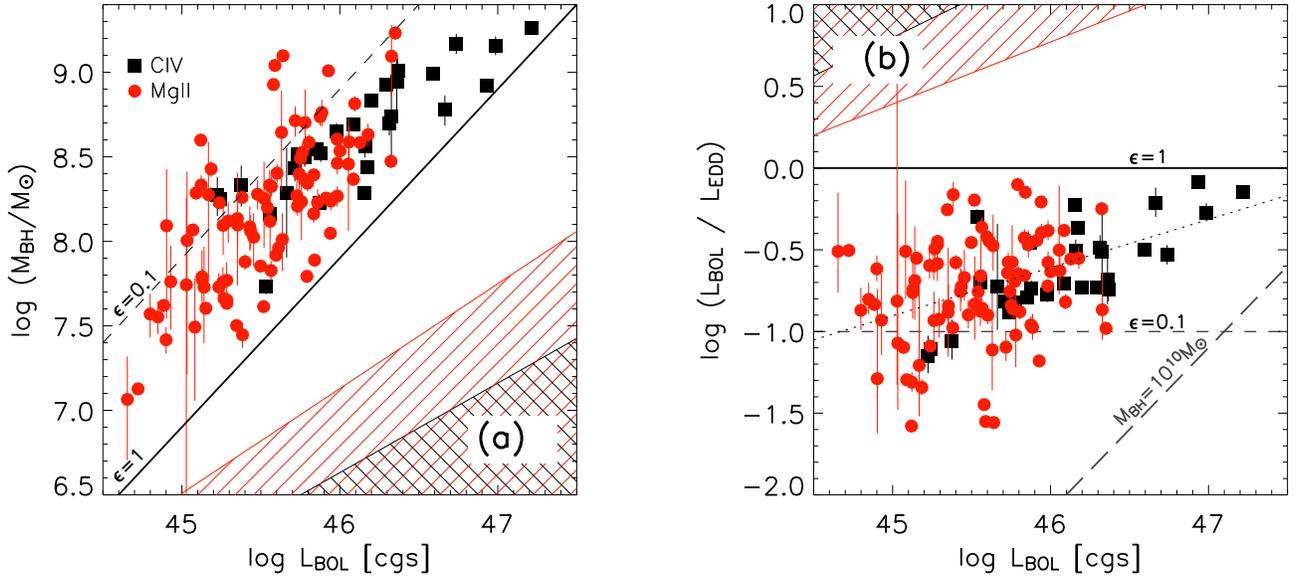}
\caption[BH mass and Eddington ratio as a function of the luminosity]
{\label{fig:MbhLedd=f(Lbol)}
  Distribution of inferred BH masses (left) and Eddington ratios (right) versus
  AGN bolometric luminosities for the VVDS sample. 
  The solid line and the dashed line correspond to Eddington ratios of 
  $\epsilon = 1$ and $\epsilon = 0.1$, respectively. 
  Different symbols denote the emission line used for the mass estimate:
  Red filled circles indicate that the black hole masses were derived from 
  \ion{Mg}{ii}, while the black squares correspond to \ion{C}{iv}.
  Error bars correspond to our uncertainties on the line width measurements.
  Inside the hashed regions, AGN would have emission lines with 
  FWHM $<$ 1000~\kms\ implying that they would have been missed in our
  sample. The dotted line in panel b shows a linear regression relation. 
  \emph{[See the online edition of the Journal for a color version of this figure.]}}
  \end{center}
\end{figure*}

The distribution of the inferred BH masses versus bolometric luminosities is
shown Fig.~\ref{fig:MbhLedd=f(Lbol)}a. As expected, there is a trend of
$M_\mathrm{BH}$ increasing with $L_\mathrm{bol}$. The overall mean and
associated error of the BH masses for the full sample is $\log M_\mathrm{BH} =
8.28 \pm 0.04$.
We split our sample at $\log L = 45.7$ into a `low luminosity' and a
`high luminosity' subset, containing respectively 62 and 58 objects.
The corresponding mean BH masses are $8.00 \pm 0.05$ and
$8.57 \pm 0.04$, respectively.   
However, the trend is not consistent with the
assumption of $L_\mathrm{bol} \propto M_\mathrm{BH}$, i.e.\ with an Eddington
ratio $\epsilon$ independent of AGN luminosity.  This is highlighted in
Fig.~\ref{fig:MbhLedd=f(Lbol)}b, where we plot $\epsilon$ versus
$L_\mathrm{bol}$ for the same objects.  
The mean $\log\epsilon$ for the full sample is $-0.71 \pm 0.03$
and has a dispersion of 0.33 dex. 
For the `low luminosity' sample the mean is $\log\epsilon = -0.81 \pm 0.04$ , 
and for the `high luminosity' subset  $\log\epsilon = -0.61 \pm 0.03$ .

The \emph{dispersion} of $\epsilon$ differs even more strongly between 
low and high luminosity subsets:
for $\log L_\mathrm{bol} > 45.7$, there is little spread in $\epsilon$
(1st and 3rd quartiles in $\log\epsilon$ are $-0.79$ and $-0.44$).
For $\log L_\mathrm{bol} < 45.7$, however, the spread is larger, with 1st
and 3rd quartiles being $-1.06$ and $-0.55$, respectively. 
A similar behavior is observed for other
percentiles. A Kolmogorov-Smirnov test comparing the distribution of 
$\epsilon$ in the two subsets gives a probability of only 1.8~\% that both
subsets were drawn from the same parent population; thus, the two subsets have
significantly different distributions in their Eddington ratios.

Most of the difference between the two $\epsilon$ distributions is due to
  the existence of a significant tail of low $\epsilon$ values for the low
luminosity AGN. In fact, the fraction of slowly accreting black holes with
$\epsilon < 0.1$ for the AGN with $\log L_\mathrm{bol} < 45.7$ (16/62) is five
times larger than the same fraction for those with $\log L_\mathrm{bol} >
45.7$ (3/58). The significance of the difference in this tail, derived on the
basis of a Fisher exact test on a  $2 \times 2$ contingency table, is at about
the 3$\sigma$ level.

The large number of low-$\epsilon$ AGN at low luminosities produces an
apparent trend of $\epsilon$ increasing with $L_\mathrm{bol}$. A formal
regression gives $\log\epsilon =  - 0.89 + 0.30 \, (\log L_\mathrm{bol} - 45)$.
We caution however against an overinterpretation of that
trend, as our sample covers only a limited range of luminosities. 
A much wider luminosity range would be needed to establish a robust
$\epsilon(L_\mathrm{bol})$ relation
 (but see the discussion in Sect.~\ref{sec:comparison} and Fig.~\ref{fig:Netzer}). 
Moreover, as also discussed below, the slope of this relations depends on
  the choice of the exponent of the empirical luminosity-size relation adopted
  in the virial scaling relations. 
The linear-Pearson ($r$) and Spearman-rank ($\rho$)
  correlation coefficients between $\log L_\mathrm{bol}$ and $\log \epsilon$
  taken alone indicate a mild correlation ($r=0.40$
  and $\rho=0.37$, respectively).

About 75\% of the AGN in our sample belong to the low-redshift, \ion{Mg}{ii}
subsample at an average redshift of $\sim 1.5$.
Since higher redshift AGN in the sample have, on average,
higher luminosities, the \ion{C}{iv} sample is populating mostly
the `high luminosity' region of Fig.~\ref{fig:MbhLedd=f(Lbol)}.
In the overlapping luminosity range ($45.5 \lesssim  \log L_\mathrm{bol}
\lesssim 46.4$), the AGN in the two redshift intervals have similar
mean Eddington ratios or BH masses. However the high redshift objects seem to
follow a steeper and tighter $\log \epsilon = \alpha \log L_\mathrm{bol} +
\mathrm{const.}$ relation ($\alpha = 0.40$, $r=0.75$)
than the low redshift sample ($\alpha = 0.22$, $r=0.24$). 
Interestingly, the difference in the best fit slope
$\alpha$ between the two subsamples disappears if one adopts scaling relations
with the same size-luminosity exponent $\gamma$. 
However, also in this case, the correlation would have a smaller scatter for
the \ion{C}{iv} AGN than for the \ion{Mg}{ii} AGN.

We now consider possible sources of systematic errors, starting with sample
incompleteness. Obviously, a selection bias against low mass black holes with
high Eddington ratios would depopulate the lower left part of the left panel in
Fig.~\ref{fig:MbhLedd=f(Lbol)}, where AGN with high $\epsilon$ would be
located. AGN with low $M_\mathrm{BH}$ and high $\epsilon$ are
characterized by relatively narrow emission lines.  As the VVDS AGN sample is
defined through the detection of broad emission lines in low-resolution
spectra, such a selection bias can in principle exist. 
However, from the spectral resolution of 350~\kms\ 
we expect the sample to be reasonably complete 
for lines intrinsically broader than $\sim 1000$~\kms.
This is shown in Fig.~\ref{fig:MbhLedd=f(Lbol)} where the areas of
incompleteness corresponding to $FWHM \le 1000$~\kms are marked as hashed
regions.
It is clear that the lack of high $\epsilon$ objects among the low luminosity
AGN cannot predominantly be due to limited spectral resolution. 

\begin{figure}
  \begin{center}
    \includegraphics[width=0.95\linewidth]{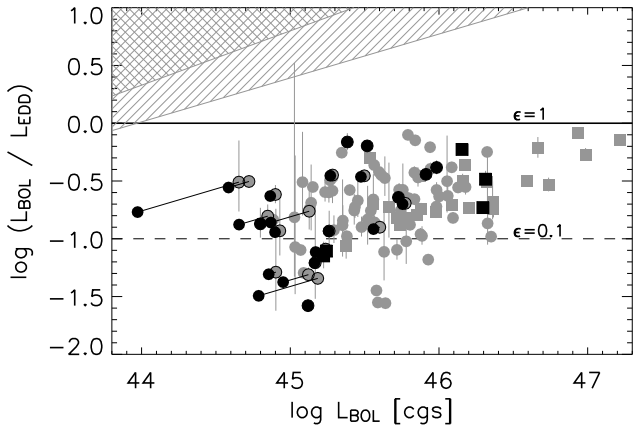} 
    \caption[Effect of host galaxy contamination]
{\label{fig:host} Effect of host galaxy contamination on the distribution
    of AGN Eddington ratios versus bolometric luminosities. All points of
    Fig. \ref{fig:MbhLedd=f(Lbol)} are reported in grey with the same symbol
    convention.  Objects for which an estimate of their host galaxy
    contamination is available are shown with an open black symbol linked to a
    filled black symbol, corresponding to the position of
    these objects, respectively before and after correction.}
\end{center}
\end{figure}

Since we are probing the AGN population down to low luminosities, host galaxy
contamination could cause us to over-estimate AGN continuum luminosities. This
would lead to an overestimation of BH masses ($M_\mathrm{BH} \propto
L^{\gamma}$) as well as Eddington ratios ($\epsilon \propto L^{1-\gamma}$).
We use here the result of the SED analysis presented in \cite{Bongiorno2007}
to estimate the host galaxy contribution to the total continuum flux. The
multi-wavelength coverage necessary for this analysis is available for
about a fourth of the objects of our sample.  Most of them (23/28)
are in the low redshift range.  Fig. \ref{fig:host} is a version of
Fig. ~\ref{fig:MbhLedd=f(Lbol)}b, corrected for this effect.  We find that
host contamination at 1500 \AA\ is negligible for all our objects. At 3000
\AA, this correction exceeds 0.1 dex in term of bolometric luminosities for 6
out of 23 AGN. These six objects are all in our 'low luminosity' sample and
therefore our conclusions are reinforced: $\sim 30$\% of the low luminosity
AGN are likely to have somewhat smaller Eddington ratios than our above
estimates.

Finally, we verified that the result presented here would have not been
significantly different if we had included also the 16 low S/N objects.

\section{Discussion}

\subsection{\label{sec:comparison}Comparison with other studies}

\citet{Kollmeier2006} determined black hole masses for a sample of $\sim 400$
AGN with optical magnitudes $R\le 21.5$, in the context of the AGES
survey. Their compilation shows a nearly constant Eddington ratio of $\sim
0.25$, with a dispersion of only $\sim 0.3$~dex, over a wide range of
luminosities and redshifts.  Our lower redshift sample overlaps with
their \ion{Mg}{ii} virial masses in the luminosity range $45 \lesssim \log
L_\mathrm{bol} \lesssim 46$ \ergs.  If we superficially compare their results
with ours in this range, we find them to be marginally inconsistent.  The
probability returned from a KS-test for the two samples to have their
Eddington ratios drawn from the same distribution is $P=9$~\% .  However,
this difference is only caused by the different recipes used to estimate
bolometric luminosities and, in particular, black hole masses.  If we
recompute the BH masses and bolometric luminosities of the
\citet{Kollmeier2006} sample with the same recipes used in the present paper,
we find that the two samples are fully consistent with each other.

In particular, \citet{Kollmeier2006} adopted a very steep exponent for the
empirical luminosity-size relation for the \ion{Mg}{ii} emission line, 
$\gamma = 0.88$ ($R\propto L^{\gamma}$), whereas we employed $\gamma =
0.62$ which is directly taken from the calibration by \cite{McLure2004}. A
larger $\gamma$ makes the $L_\mathrm{bol}$-$M_\mathrm{BH}$ relation appear
steeper and results in smaller $M_\mathrm{BH}$ and higher $\epsilon$ values 
for the lower luminosity AGN.

However, we believe that there are good reasons against such a high value of
$\gamma$. Recent reverberation mapping studies \citep{Paltani2005,Kaspi2007}
suggest a rather flat luminosity-size relation also at high luminosities with
$\gamma$ even approaching 0.5 (corresponding to an approximately
luminosity-independent ionization parameter in the broad-line region of
AGN). A low value of $\gamma \sim 0.5$ is also indicated for low redshift AGN
after correction for host galaxy contributions \citep{Bentz2006}. Our adopted
value of $\gamma = 0.62$ may therefore even be considered conservative.

\begin{figure}
  \begin{center}
    \includegraphics[width=0.95\linewidth]{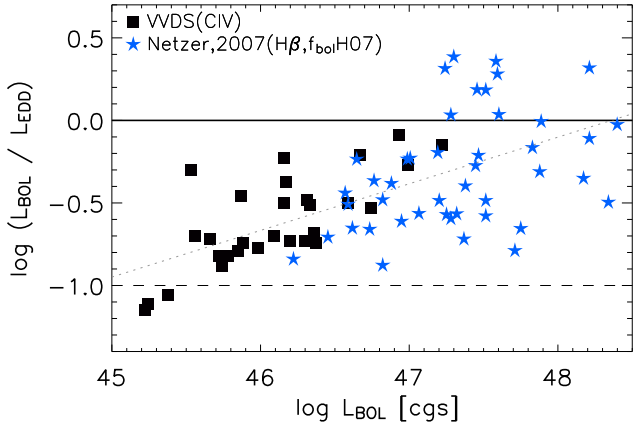}
    \caption[Comparison with Netzer (2007)]
{\label{fig:Netzer} Distribution of Eddington ratios versus
    bolometric luminosities of VVDS high redshift and \cite{Netzer2007}
    sample. For comparison purpose, Eddington ratios and bolometric
    luminosities of the latter sample have been recomputed with a
    luminosity dependent correction factor ($9.6 < f_\mathrm{bol}(5100\mbox{\AA})
    < 10.5$, \citet{Hopkins2007}). The dotted line shows a linear regression
    relation ($\log\epsilon \propto 0.29 \log L_\mathrm{bol}$).}
\end{center}
\end{figure}

More recently, \cite{Netzer2007} also found a positive trend of $\epsilon$ with
luminosity for AGN with redshift 2.3--3.4.  

They derived virial black hole masses from the redshifted
\ion{H}{$\beta$} line observed in the near infrared, thus applying directly
the reverberation mapping based calibration (although some
extrapolation towards high luminosities was required).

Given the match in redshift range, we decided to combine the results by Netzer
et al.\ with our \ion{C}{iv} sample.  The results are shown in
Fig.~\ref{fig:Netzer} where, for consistency, we have recomputed the
bolometric luminosities and Eddington ratios of the AGN in Netzer's sample
using the same bolometric corrections we employed for our data, i.e.\
applying the \cite{Hopkins2007} luminosity-dependent SED model.  The combined
sample covers now more than three orders of magnitude in luminosity, with most
of our objects being much fainter than those in Netzer's sample. For the
luminosity range common to both samples, the values of $\epsilon$ are in very
good agreement, despite the fact that we could only use the less trustworthy
\ion{C}{iv} lines.

Looking at the combined dataset there is again evidence for a correlation of
Eddington ratios. In fact, the best-fit regression (dotted line in
Fig.~\ref{fig:Netzer}) returns $\log \epsilon = -0.97 + 0.28\,\log
(L_\mathrm{bol} - 45)$, which is almost indistinguishable from the fit to only
the combined VVDS \ion{Mg}{ii} + \ion{C}{iv} sample. Thus the slow increase of
Eddington ratios with bolometric luminosities seems to be a remarkably
continuous property of high-redshift type~1 AGN, from the very low
luminosities of the VVDS AGN to the highly luminous quasars in the sample of
Netzer et al.  (On the other hand, the dispersion of Eddington ratios in the
Netzer sample is larger than in the VVDS, in particular due to the significant
number of super-Eddington objects in the former.)

Very recently, \citet{Shen2008} employed a very large sample of SDSS quasar
spectra to investigate systematic effects and biases in the derivation of
relations between luminosities and black hole masses. They essentially confirm
the low apparent dispersion in $\epsilon$ of $\la 0.3$ found already by
\citet{Kollmeier2006}, however with the exception of their lowest luminosity (and
also lowest redshift, $z < 1$) bin where the dispersion increases to
0.42~dex. Thus, while there is essentially no overlap in the
luminosity-redshift plane between SDSS and VVDS, the trends observed in our
VVDS sample seems to be consistent with the SDSS results.  

\citet{Babic2007} argue that an apparent trend of $\epsilon$ as a function of
luminosity is expected if one convolves a double power-law black hole mass
function with a relatively broad distribution of Eddington ratios truncated at
$\epsilon = 1$. We note however that the observed distribution of $\epsilon$
is too symmetric and too narrow for this to be a strong effect: The apparent
upper bound of $\epsilon$ evolves along with $L_\mathrm{bol}$ from $\log
\epsilon < - 0.5$ at $\log L_\mathrm{bol} \sim 45$ to $\log \epsilon <
0.5$ for the high-luminosity objects in the sample by \citet{Netzer2007}. In
other words, 
there is no clear evidence for a physical truncation at a fixed $\epsilon$.
It is of course still possible that AGN accretion physics imposes some unknown
biases on the distribution of Eddington ratios, which may even depend on
luminosity or black hole mass, in which case an effect such as described by
\citet{Babic2007} may become relevant at some level. Much larger samples 
and a better understanding of the underlying physical processes would be 
required to investigate such effects.

\subsection{Relation to the AGN luminosity function}

At fixed redshift, the AGN luminosity function (AGNLF) is generally described
as a double power-law. It has now become clear that its shape evolves with
redshift, with a marked break for $z>1$ which almost disappears at lower
redshift, as the faint-end slope steepens towards later cosmic times
\citep[e.g.][]{Hasinger2005,Hopkins2007,Bongiorno2007}.  The luminous part of
the AGNLF is dominated by black holes that appear to be typically accreting
close to the Eddington limit ($\epsilon \sim 0.1$--1), with relatively little
dispersion, so that luminosities are roughly proportional to black hole
masses, and this part of the AGNLF closely mirrors the black hole mass
function.

The flat part of the AGNLF, on the other hand, could be composed of either low
mass black holes also accreting close to Eddington, or of high-mass black
holes with very low accretion rates, or of a mixture.  In the context of a
simple model where black hole growth and nuclear activity is triggered by
galaxy mergers, \citet{Cattaneo2001} first suggested that the faint end slope
of the AGNLF could be dominated by objects observed in the decaying phase of
their light curve, well past their peak of activity.  This idea was recently
followed up by numerical simulations of galaxy mergers incorporating AGN
feedback.  For example \cite{Hopkins2006b} find that the observed redshift
evolution of the faint-end slope of the AGN luminosity function (flatter at
higher redshift) is well reproduced with the luminosity-dependent quasar
lifetime that they derive from extensive numerical simulation of galaxy
mergers.

Our observations show that while some of the low-luminosity AGN in our sample
have just low $M_\mathrm{BH}$, many have instead the properties (i.e. high
$M_\mathrm{BH}$, low $\epsilon$) predicted by these models. This is consistent
with the suggested picture in which the faint end of the AGN luminosity function is
populated with black holes that have exhausted a substantial fraction of their
fuel.  We speculate that at these redshifts we see glimpses of a population of
AGN with black hole masses similar to those of luminous quasars, but already
half starved and on their way to get extinguished.

From an analysis of a heterogeneous sample of low redshift AGN, \cite{Woo2002}
find that small Eddington ratios are found mainly for AGN with $\log
L_\mathrm{bol} \le 44.5$, which in their sample are represented only by local
Seyfert galaxies at $z \le 0.1$ (see their Fig.~8).  Comparing their results
with our measurements at $z\ga 1$ suggests that the luminosity below which
such small Eddington ratios are found may evolve with redshift.  This is, at
least qualitatively, consistent with the observed redshift evolution of the
break of the AGN luminosity function.

\section{Conclusions}

The VVDS is the first large spectroscopic AGN survey to probe luminosities as
low as $\log L_\mathrm{bol} \lesssim 45$ at redshifts $z > 1$. We estimated
black hole masses and Eddington ratios for 120 AGN. The main result
of our study is a marked increase of the dispersion in Eddington ratios
towards lower AGN bolometric luminosities.  A substantial fraction of black
holes in low-$L$ AGN accretes at less than 10~\% of their Eddington limits,
whereas such low accretors are rare among AGN with higher $L$. 

Our data also suggest that on average, the Eddington ratios systematically
increase with nuclear luminosity. In the presence of substantial scatter and
limited luminosity coverage, this trend is not easy to quantify; the slope of
a relation $\epsilon \propto L^\alpha$ depends on the adopted exponent in the
empirical luminosity-size relation needed for virial scaling relations.
Tentatively combining our data with those of \citet{Netzer2007}, however, leads
to fully consistent results and underlines the indicated trend of $\epsilon$
increasing with $L_\mathrm{bol}$.

It is currently widely discussed how accurate the black hole masses and
Eddington ratios based on single-epoch spectra can be. The best line is
clearly H$\beta$ as here the luminosity-size relation has been directly
calibrated with reverberation mapping. \ion{Mg}{ii}-based estimates can be
cross-calibrated with H$\beta$ measurements and correlate quite well
\citep{McLure2002,Shen2008}.  The \ion{C}{iv} line, on the other hand, is under
suspicion of representing gas that is not necessarily in or even close to
virial equilibrium.  One strong indication for such non-gravitational effects
is the systematic blueshift of \ion{C}{iv} with respect to low ionization
lines \citep{Gaskell1982,Tytler1992}, which in combination with often
asymmetric profiles \citep{Richards2002} can be interpreted as the
result of 
obscuration or radiative pressure. Consequently, the \ion{C}{iv} 
emission line is often considered as not well suited to estimate black hole 
masses. \cite{Baskin2005} and \cite{Netzer2007} found only a weak correlation
between virial black hole mass estimates based on H$\beta$ and
\ion{C}{iv}. Similarly, \cite{Shen2008} noted a much tighter correlation 
between H$\beta$ and \ion{Mg}{ii} than between \ion{Mg}{ii} and 
\ion{C}{iv}. For our VVDS sample, however, the observed trends between
\ion{Mg}{ii} and \ion{C}{iv} based subsamples (and also the H$\beta$ sample
by \citealt{Netzer2007}) are highly consistent. In fact, the observed 
scatter in the $L_\mathrm{bol}$-$\epsilon$ relation is \emph{lower} for 
the \ion{C}{iv} objects than for the \ion{Mg}{ii} ones. It may be that
radiation pressure and outflows are relevant in particular for high-luminosity
QSOs (as has been also suggested by \citealt{Marconi2008}), and that therefore
virial mass estimates based on \ion{C}{iv} are more reliable for the faint 
AGN sampled in the VVDS than for other surveys.

While the `virial estimator' is likely to remain for some time the only
practical method to obtain statistics on black hole masses at substantial
redshifts, the present dependency of all measurements on the small number of
low-$z$ reverberation-mapped AGN is unsatisfactory. It would be highly
desirable if directly calibrated luminosity-size relations could be
established also for higher redshifts and other lines than H$\beta$.

\acknowledgements{
We thank the referee, Marco Salvati, for his  constructive
  comments that have led to improve this publication.
 We are grateful to Marianne Vestergaard for providing us with her UV
 \ion{Fe}{II} templates, as well as to Juna Kollmeier and Hagai Netzer
   for communicating 
 us the data table of their publications for our comparison purpose.
 We thank Suzy Collin and Asmus B\"ohm for helpful discussions.} 

\bibliographystyle{aa}
\bibliography{qso}

\newpage

\appendix
\section{Mbh tables}
\label{sec:tabMbh}

\begin{longtable}{lccrccrcr}
\caption{\label{tab:Mbh_MgII} Black hole masses and Eddington ratios estimated from the \ion{Mg}{ii} emission line}\\ 
  \hline\hline
  $ Object \, ID $ & $z$ & $ I_{AB} $&
  \hfill $S/N$ \hfill&$ \log{\lambda L_{3000}}$&$ \log{L_\mathrm{bol}}$&$FWHM$&
  $\log{\frac{M_{BH}}{M\sun}}$&$\log{\epsilon}$\\
  \multicolumn{1}{c}{(1)} & (2) & (3)      &\multicolumn{1}{c}{(4)}& (5)
  &(6)                            &\multicolumn{1}{c}{(7)}&(8)
  &\multicolumn{1}{r}{(9)}\\ 
  \hline
  \endfirsthead
  \caption{continued.}\\
  \hline\hline
  $ Object \, ID $ & $ \hfill   z \hfill $ & $ I_{AB} $&
  $ S/N $&$ \log{\lambda L_{3000}}$&$ \log{L_\mathrm{bol}}$&$FWHM$&
  $\log{\frac{M_{BH}}{M\sun}}$&$\log{\epsilon}$\\
  \hline
  \endhead
  \hline
  \endfoot
220551387 &   1.881 &  21.55 &   11.5 &  45.00 &  45.83 &   3277&   8.16&  -0.43 \\
140265284 &   1.855 &  21.45 &   16.7 &  44.89 &  45.73 &   3730&   8.21&  -0.58 \\
220082140 &   1.848 &  20.68 &   18.2 &  45.23 &  46.05 &   3903&   8.46&  -0.50 \\
100113463 &   1.844 &  20.52 &   34.0 &  45.27 &  46.09 &   5720&   8.81&  -0.82 \\
220090821 &   1.833 &  20.49 &   15.5 &  45.26 &  46.09 &   3442&   8.37&  -0.38 \\
140520998 &   1.829 &  20.89 &   26.5 &  45.06 &  45.89 &   6286&   8.76&  -0.98 \\
140363408 &   1.826 &  21.87 &   11.3 &  44.97 &  45.81 &   5448&   8.58&  -0.88 \\
100110223 &   1.826 &  21.35 &    7.9 &  44.92 &  45.76 &   3770&   8.23&  -0.58 \\
\end{longtable}

\begin{longtable}{lccrccrcr}
\caption{\label{tab:Mbh_CIV} Black hole masses and Eddington ratios estimated from the \ion{C}{iv} emission line}\\ 
  \hline\hline
  $ Object \, ID $        & $z$ & $I_{AB}$ &\hfill $S/N$ \hfill    &
  $\log{\lambda L_{1350}}$&$ \log{L_\mathrm{bol}}$&$\sigma$
  &$\log{\frac{M_{BH}}{M\sun}}$&$\log{\epsilon}$\\ 
  \multicolumn{1}{c}{(1)} & (2) & (3)      &\multicolumn{1}{c}{(4)}& (5)
  &(6)                            &\multicolumn{1}{c}{(7)}&(8)
  &\multicolumn{1}{r}{(9)}\\ 
  \hline
  \endfirsthead
  \caption{continued.}\\
  \hline\hline
  $ Object \, ID $        & $z$ & $I_{AB}$ &\hfill $S/N$ \hfill    & $\log{\lambda L_{1350}}$&$ \log{L_\mathrm{bol}}$&$\sigma$               &$\log{\frac{M_{BH}}{M\sun}}$&$\log{\epsilon}$\\
  \hline
  \endhead
  \hline
  \endfoot
220183694 &   4.193 &  20.27 &   33.1 &  46.34 &  46.93 &   2976&   8.92&  -0.09 \\
020254576 &   3.853 &  21.15 &   35.5 &  45.71 &  46.31 &   3406&   8.70&  -0.48 \\
220267678 &   3.833 &  21.44 &   11.9 &  45.72 &  46.33 &   3534&   8.74&  -0.51 \\
020277536 &   3.626 &  23.55 &   16.7 &  44.58 &  45.24 &   4032&   8.25&  -1.11 \\
220055529 &   3.594 &  21.52 &    9.6 &  45.76 &  46.37 &   4708&   9.01&  -0.74 \\
100471137 &   3.494 &  20.12 &   39.6 &  46.07 &  46.67 &   2994&   8.78&  -0.21 \\
020179116 &   3.308 &  23.89 &   10.4 &  44.56 &  45.22 &   4191&   8.27&  -1.15 \\
140465826 &   3.290 &  20.16 &   60.3 &  46.14 &  46.74 &   4486&   9.17&  -0.53 \\
\end{longtable}

\newpage

\section{Catalog of VVDS broad-line AGN (first and second-epoch data)}
\label{epoch2}
\begin{longtable}{lrrrrccccccc}
  \caption{\label{tab:AB} Catalog of broad-line AGN with secure redshift.}   \\
  \hline\hline
  $Object\,ID$
  &\multicolumn{1}{c}{$\alpha_{J2000}$}&\multicolumn{1}{c}{$\delta_{J2000}$}&\multicolumn{1}{c}{$z$}
  &$Flag$&$Epoch$&$ B_{AB} $&$ V_{AB} $&$ R_{AB} $&$ I_{AB} $&$Morphology
  $&$Remark$\\
  \multicolumn{1}{c}{(1)} &\multicolumn{1}{c}{(2)}             &
  \multicolumn{1}{c}{(3)}            & \multicolumn{1}{c}{(4)}     &  (5)   &
  (6)  &  (7)     &  (8)     &  (9)     &  (10) & (11) & (12)\\
  \hline
  \endfirsthead
  \caption{Catalog of broad-line AGN with secure redshift (continued).}\\
  \hline\hline
  $ Object \, ID $&$ \hfill  \alpha_{J2000} \hfill $&$ \hfill  \delta_{J2000} \hfill $&$ \hfill   z \hfill $&$Flag$&$Epoch$&$ I_{AB} $&$ B_{AB} $&$ V_{AB} $&$ R_{AB} $&$Morphology      $&$Remark$\\
  \hline
  \endhead
  \hline
  \endfoot
 CDFS: & deep \, mode &          10\, AGN \hfill & & & \\ \hline
000018607 & $03^{h} 32^{m} 18.26^{s} $ & $-27^{o} 52^{m} 41.42^{s} $ & $ 2.8010$ & $ 14$ & $1$ & $23.94$ & $24.62$ & $24.08$ & $23.91$ & \dots          & \dots \\
000025363 & $03^{h} 32^{m} 59.85^{s} $ & $-27^{o} 47^{m} 48.42^{s} $ & $ 2.5673$ & $ 14$ & $1$ & $22.09$ & $21.66$ & $21.92$ & $21.95$ & \dots          & \dots \\
000023526 & $03^{h} 32^{m} 43.25^{s} $ & $-27^{o} 49^{m} 14.38^{s} $ & $ 1.9199$ & $ 14$ & $1$ & $22.53$ & $22.70$ & $22.65$ & $22.17$ & \dots         & Sz. \\
000033629 & $03^{h} 32^{m} 25.17^{s} $ & $-27^{o} 42^{m} 19.05^{s} $ & $ 1.6207$ & $ 14$ & $1$ & $24.33$ & $23.90$ & $22.94$ & $22.14$ & \dots          & \dots \\
000073509 & $03^{h} 32^{m} 02.47^{s} $ & $-27^{o} 46^{m} 00.53^{s} $ & $ 1.6199$ & $ 14$ & $1$ & $25.34$ & $24.59$ & $23.95$ & $23.62$ & \dots         & Sz. \\
000028880 & $03^{h} 33^{m} 03.62^{s} $ & $-27^{o} 45^{m} 18.97^{s} $ & $ 1.2574$ & $ 14$ & $1$ & $23.78$ & $23.30$ & $22.98$ & $22.71$ & \dots         & Sz. \\
000031947 & $03^{h} 32^{m} 00.37^{s} $ & $-27^{o} 43^{m} 19.85^{s} $ & $ 1.0401$ & $ 14$ & $1$ & $22.99$ & $22.25$ & $22.13$ & $22.12$ & \dots         & Sz. \\
\end{longtable}

\begin{table}
\caption[]{\label{tab:C} AGN with a single emission line detected (flag 19).}

\resizebox{0.85\linewidth}{!}{
\begin{tabular}{lrrcrlccccc}
  \hline\hline
  $Object\,ID$            &\multicolumn{1}{c}{$\alpha_{J2000}$}&\multicolumn{1}{c}{$\delta_{J2000}$}&$Epoch$&\multicolumn{1}{c}{$\lambda_{BL}$}&\multicolumn{1}{c}{$ z \, solutions $}&$ I_{AB} $&$  B_{AB} $&$ V_{AB} $&$ R_{AB} $&$Morphology$\\
  \multicolumn{1}{c}{(1)} & \multicolumn{1}{c}{(2)}            & \multicolumn{1}{c}{(3)}            & (4)   &\multicolumn{1}{c}{(5)}           &\multicolumn{1}{c}{  (6) }            &  (7)     &  (8)     &  (9)     &  (10) & (11) \\
  \hline
CDFS: & deep\, mode  \hfill&           2\, AGN \hfill &&&&&&&\\ \hline
000031270 &$ 03^{h} 32^{m} 57.74^{s}  $&$ -27^{o} 43^{m} 50.12^{s} $ & $1 $&$ 7319. $&$ \,  0.1150^a\,  1.6150^b  $ & $23.81$ & $23.65$ & $23.44$ & $23.52$ & \dots          \\
000017025 &$ 03^{h} 31^{m} 54.30^{s}  $&$ -27^{o} 53^{m} 49.58^{s} $ & $1 $&$ 6500. $&$  \,  1.3220^b\,  2.4050^c\,  3.1960^d$ & $24.78$ & $24.51$ & $23.93$ & $23.70$ & \dots          \\
\hline
0226-04 : & deep\, mode  \hfill&           6\, AGN \hfill &&&&&&&\\ \hline
020375508 &$ 02^{h} 25^{m} 48.99^{s}  $&$ -04^{o} 10^{m} 28.04^{s} $ & $1 $&$ 5963. $&$  \,  1.1300^b\,  2.1240^c\,  2.8500^d$ & $23.25$ & $22.98$ & $23.05$ & $22.54$ & \dots          \\
020293248 &$ 02^{h} 26^{m} 25.92^{s}  $&$ -04^{o} 21^{m} 12.73^{s} $ & $1 $&$ 7335. $&$ \,  0.1180^a\,  1.6210^b  $ & $24.79$ & $24.54$ & $24.14$ & $23.26$ & \dots          \\
\end{tabular}
}
\end{table}
\end{document}